\documentclass[conference]{IEEEtran}
\pdfoutput=1

\usepackage{graphicx}
\usepackage{wrapfig}
\usepackage{subfigure}
\usepackage{stfloats}
\usepackage{hyperref}
\usepackage{url}
\usepackage{amsmath}
\usepackage{footnote}
\makesavenoteenv{tabular}
\makesavenoteenv{table}
\usepackage{multirow}
\usepackage{fixltx2e}

\usepackage[usenames, dvipsnames]{color}
\newcommand{\diff}{\small \texttt{diff}}
\newcommand{\code}[1]{\texttt{#1}}

\begin{document}
\title{Automatically Generating Commit Messages \\ from Diffs using Neural Machine Translation}

\author{\IEEEauthorblockN{ Siyuan Jiang,
Ameer Armaly, and
Collin McMillan}

\IEEEauthorblockA{Department of Computer Science and Engineering\\
University of Notre Dame,
Notre Dame, IN, USA\\ Email: \{sjiang1, aarmaly, cmc\}@nd.edu}
}


\maketitle

\begin{abstract}
Commit messages are a valuable resource in comprehension of software evolution, since they provide a record of changes such as feature additions and bug repairs.  Unfortunately, programmers often neglect to write good commit messages.  Different techniques have been proposed to help programmers by automatically writing these messages.  These techniques are effective at describing what changed, but are often verbose and lack context for understanding the rationale behind a change.  In contrast, humans write messages that are short and summarize the high level rationale.  In this paper, we adapt Neural Machine Translation (NMT) to automatically ``translate'' diffs into commit messages.  We trained an NMT algorithm using a corpus of diffs and human-written commit messages from the top 1k Github projects.  We designed a filter to help ensure that we only trained the algorithm on higher-quality commit messages.  Our evaluation uncovered a pattern in which the messages we generate tend to be either very high or very low quality.  Therefore, we created a quality-assurance filter to detect cases in which we are unable to produce good messages, and return a warning instead.
\end{abstract}

\section{Introduction}
\vspace{-0.2cm}

Commit messages are natural language descriptions of changes in source code.  When a programmer updates code, a typical procedure is to upload the change to a version control system with a short commit message to describe the purpose of the change, e.g., ``adds support for 9 inch tablet screen size.''  The repository stores the message alongside a \texttt{diff} that represents the difference between the current and previous version of the affected files.  The practice is extremely common: for this paper alone, we obtained over 2M diffs and messages from just 1k projects on GitHub.

Commit messages are useful because they help programmers to understand the high level rationale for a change without reading the low level implementation details.  They serve a valuable purpose in comprehension of software evolution, and act as a record of feature additions and bug repairs~\cite{Buse:2010:ADP:1858996.1859005}.  Unfortunately, programmers sometimes neglect commit messages~\cite{6606588, 5463344}, likely due to the same time and market pressures that have been reported to affect many types of documentation~\cite{Roehm:2012:PDC:2337223.2337254, Fluri:2007:CCC:1339262.1339530, Kajko-Mattsson:2005:SDP:1032622.1035374}.  In short, programmers use commit messages but often avoid writing them themselves.

Automated generation of commit messages has been proposed as an alternative to manual efforts by programmers.  For example, Buse~\emph{et al.}~\cite{Buse:2010:ADP:1858996.1859005} describe DeltaDoc, a tool that summarizes what changed in the control flow of a program between code versions.  Likewise, Cortes-Coy~\emph{et al.}~\cite{Linares2015ICSE} built ChangeScribe, which summarizes changes such as method additions.  These and other existing techniques (see Section~\ref{sec:related_work}) have been shown to be effective in answering questions about what changed and where from one code version to another.

What is missing from existing approaches is a short, high level description of the purpose behind commits.  Current approaches are effective at summarizing what changed and where, but do not answer the question \emph{why}~\cite{Buse:2010:ADP:1858996.1859005}.  Questions of \emph{why} traditionally require human insight since they involve synthesis of different, complex data sources and context.  However, as Mockus~\emph{et al.}~\cite{883028} observed, many commit messages are similar and can be broadly categorized as related to bug repair, feature additions, etc.  Plus, they follow similar grammatical patterns such as verb-direct object structure (e.g. ``adds support for...'')~\cite{jiang2017icpc-era}.  This observation leads us to believe that the text of commit messages can be learned and predicted if there is sufficient data.  Our view is in line with the hypothesis of ``naturalness'' of software~\cite{Hindle:2012:NS:2337223.2337322}, that software artifacts follow patterns that can be learned from sufficiently large datasets.

In this paper, we adapt a neural machine translation (NMT) algorithm to the problem of commit message generation.  Several NMT algorithms have been designed to translate between natural languages by training a neural network on pairs of sentences that humans have already translated.  The datasets required are enormous by typical software engineering research standards, involving up to tens of millions of pairs of sentences~\cite{Sennrich2016WMT,Luong2016CoRR}.  We trained an NMT algorithm using pairs of \texttt{diff}s and commit messages from 1k popular projects on GitHub.  While we were able to obtain quite large datasets (over 2M commits), we encountered many commit messages that were gibberish or very low quality (a problem others have observed~\cite{6606588, 5463344}), which if left in the training data could be reflected in the NMT's output. Therefore, we designed a filter to ensure that we only trained the algorithm using messages with a verb-direct object pattern. 

We investigate and report the effectiveness of the predictions from the process.  We found promising results as well as key constraints on the accuracy of the predictions.  In short, the NMT process performed quite well under select conditions, but poorly in others.  We report these results and promising and poor conditions as a guide to other researchers and platform for advancement in this research area.  To further promote advancement of the area, we make our implementation and data freely available in an online replication package.

Our approach has two key advantages that make it a supplement to, rather than a competitor of, existing automatic commit message generation techniques.  First, we produce short summary messages rather than exhaustive descriptions of code changes.  And second, our approach produces messages for changes to many types of software artifact in the repository, not solely source code.

\subsection{The Problem}
\label{sec:problem}
In this paper we target the problem of automatically generating commit messages.  Commit messages are useful in the long term for program comprehension and maintainability, but cost significant time and effort in the short term.  These short term pressures lead programmers to neglect writing commit messages, like other types of documentation~\cite{6606588, 5463344, Roehm:2012:PDC:2337223.2337254, Fluri:2007:CCC:1339262.1339530, Kajko-Mattsson:2005:SDP:1032622.1035374}.  Buse~\emph{et al.}~\cite{Buse:2010:ADP:1858996.1859005} point out that programmers use commit messages for two reasons: 1) to summarize \emph{what} changed, and 2) to briefly explain \emph{why} the change was necessary.  To date, research into commit message generation has exclusively focused on the question \emph{what}.  In this paper, we seek to begin answering \emph{why}.

Existing commit message generation techniques produce relatively long messages that include details such as the methods that were added or the number of files changes (\emph{what} information).  While useful, these techniques are a complement to, rather than a replacement for, high level \emph{why} information that humans write such as ``adds support for 9 inch tablet screens.''  Normally, this high level information requires human judgment.  But we hypothesize that there are patterns of commits, and that these patterns can be detected and used to generate messages for similar commits later.  Given a large number of pairs of \texttt{diff}s and messages, we believe we can train an algorithm to write new messages for new commits, based on the new commits' similarity to older ones.

\textbf{Please note} that we do not claim to generate new insights for completely new types of commits -- that task is likely to remain in the hands of human experts.  However, we do aim to write messages that reflect knowledge that can be learned from records of previous commits.  In the long run, we hope that this technology will help reduce manual effort by programmers in reading and understanding code changes in repositories.

\subsection{Paper Overview}

\begin{figure}[!b]
\centering
\includegraphics[width=2.3in]{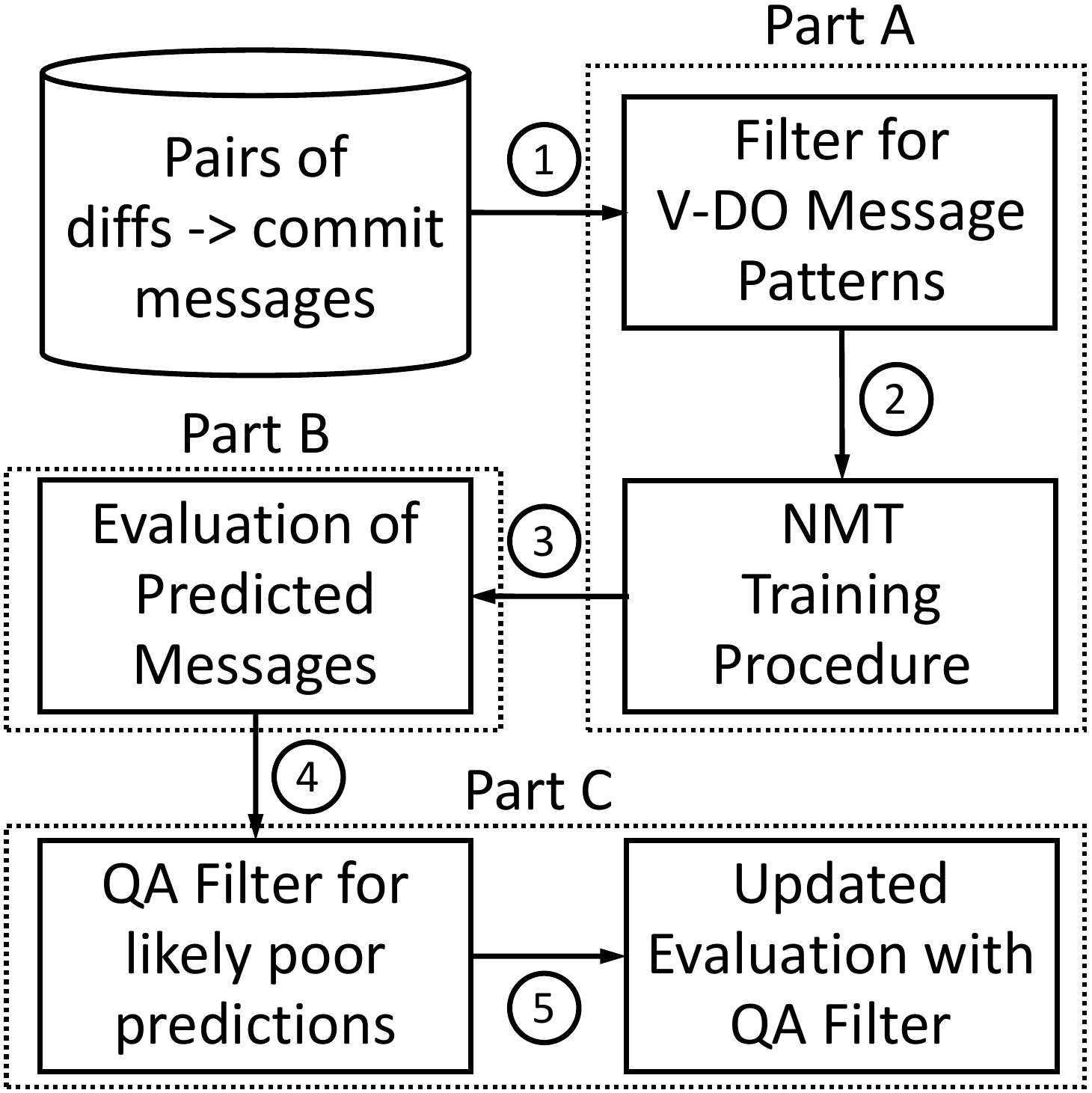}
\caption{The overview of our paper.}
\label{fig:paper_overview}
\end{figure}

Figure~\ref{fig:paper_overview} depicts an overview of our paper.  We have divided the work into three segments: In Part A (Section~\ref{sec:approach}), we present our approach to filtering for verb/direct-object (V-DO) commit message patterns and training an NMT algorithm to produce messages with this pattern.  The V-DO filter was introduced because a large percentage of the messages in the repositories we downloaded were very low quality, and we needed to ensure that we trained the NMT algorithm only with examples matching an acceptable pattern. We then trained an NMT algorithm on the pairs of \texttt{diffs} and commit messages where the messages followed the V-DO pattern.

In Part B (Sections~\ref{sec:eval} and~\ref{sec:human_study}), we evaluate the quality of the commit messages produced by the algorithm with an automated method and a human study with 2 Ph.D. students and 18 professional programmers.  We observe that while there are a significant number of positive results, there are also a significant number of negative results.  Therefore, in Part C (Sections~\ref{sec:qa_filter}), we design a quality assurance (QA) filter to detect cases in which the NMT algorithm is likely to produce a negative result.  We then modify our approach to produce a warning message instead of a commit message in those cases, and update our evaluation to show the effects of our modification.  In short, we reduce the number of poor predicted messages by 44\% at a cost of also mistakenly reducing high quality predictions by 11\%.

\section{Related Work}
\label{sec:related_work}
We split the related work into three categories:  1) the work that generates commit messages; 2) the work that summarizes source code; and 3) the work that applies deep learning algorithms in software engineering. 

\subsection{Commit Message Generation Techniques}
\label{subsec:commit_msg_tech}
We categorize the commit message generation techniques into three groups based on the inputs of the techniques. \textbf{The first group} uses code changes of a commit as an input, and summarizes the changes to generate the commit message. For example, Buse \emph{et al.} have built DeltaDoc, which extracts path predicates of changed statements, and follows a set of predefined rules to generate a summary~\cite{Buse:2010:ADP:1858996.1859005}. Similarly, Linares-V\'{a}squez \emph{et al.} have built ChangeScribe, which extracts changes between two Abstract Syntax Trees and summarizes the changes based on predefined rules~\cite{Linares2015ICSE}.


Supplementing the first group, \textbf{the second group} is based on related software documents. For example, Le \emph{et al.} have built RCLinker, which links a bug report to the corresponding commit message~\cite{Le2015ICPC}. Rastkar and Murphy have proposed to summarize multiple related documents for commits~\cite{Rastkar2013ICSE}. Integrating the ideas of the first and the second groups, Moreno \emph{et al.} have built ARENA, which summarizes changes and finds related issues to generate release notes~\cite{Moreno2017TSE}. 


\textbf{The third group} is our technique using {\diff}s (generated by ``{\small \texttt{git diff}}'') as inputs. Our technique is to translate a \texttt{diff} to a natural language sentence. Our technique supplements the first group in two ways. First, the techniques in the first group often generate multi-line summaries that contain pseudocode and template text. In contrast, our technique generates one-sentence descriptions, which can be used as a headline of the multi-line summaries. Second, our technique summarizes both code and non-code changes in {\diff}s. 


\subsection{Source Code Summarization}
Source code summarization techniques generate descriptions of source code pieces. The algorithms of the techniques can be adapted to generate summaries for changes in commits. Code summarization can be categorized into two groups: extractive and abstractive. Extractive summarization extracts relevant parts of source code and uses the relevant parts as a summary~\cite{Haiduc:2010:UAT:1919284.1919577}. Abstractive summarization includes information that is not explicitly in the source code. For example, Sridhara \emph{et al.} has designed a Natural Language Generation (NLG) system to create summaries of Java methods~\cite{Sridhara:2010:TAG:1858996.1859006}. First, the NLG system finds important statements of a Java method. Second, the system uses a text generation algorithm to transform a statement to a natural language description. This algorithm has predefined text templates for different statement types, such as {\small \texttt{return}} statements and {\small \texttt{assignment}} statements. Both DeltaDoc and ChangeScribe (discussed in Section~\ref{subsec:commit_msg_tech}) follow the similar NLG design.

Besides the NLG approach to generate abstractive summaries, Iyer \emph{et al.} have built Code-NN, which uses an Neural Machine Translation (NMT) algorithm to summarize code snippets~\cite{iyer2016acl}. This work is similar to our technique because our technique also uses an NMT algorithm. There are two key differences between our technique and Code-NN. First, the goal of Code-NN is summarizing code snippets and the goal of our technique is summarizing changes. Second, Code-NN parses code snippets and removed all the comments. In contrast, our technique's input is an entire {\diff} with code, comments, and {\diff} marks (e.g., \texttt{+} denoting insertion).

\subsection{Deep Learning in Software Engineering}
Deep learning algorithms are becoming more prevalent in Software Engineering research. Deep learning algorithms, as applied to software, automatically learn representations of software artifacts. For example, to detect code clones, traditional approaches predefine the representations of code fragments (some techniques use token sequences to represent code~\cite{Kamiya2002TSE}; others use graphs~\cite{Krinke2001WCRE,Chen2014ICSE}). In contrast, the deep learning approach introduced by White \emph{et al.}~\cite{White2016ASE} learns the representations of code automatically. Similarly, deep learning algorithms are introduced in bug localization~\cite{Lam2015ASE}, software traceability~\cite{Guo2017ICSE}, and code suggestions~\cite{White2015MSR}. 


Our technique is similar to the work done by Gu \emph{et al.}~\cite{Gu2016FSE}, because both our and their techniques use Neural Machine Translation (NMT).
Gu \emph{et al.} use NMT to translate natural language queries to API method sequences~\cite{Gu2016FSE}. Similarly, several code generation techniques use NMT to translate natural language to programming language~\cite{Ling2016ACL,Yin2017ACL}. In contrast, our technique translates {\diff}s to natural language. 

Our technique is also similar to the work by Alexandru \emph{et al.}~\cite{Alexandru2017ICPC-ERA}, which investigates the suitability of NMT for program comprehension. Alexandru \emph{et al.} use NMT for source code tokenization and token annotation. While Alexandru \emph{et al.} target on lower-level source code understanding (token-level), we target on understanding higher-level of mixtures of code and text ({\diff}-level).


\section{Background}
\label{sec:background}
We split the background section into three subsections. The first subsection is about the empirical studies on commit messages, which motivate us to generate short descriptions of commits. The second subsection describes \emph{RNN Encoder-Decoder}, a popular Neural Network Translation model, which is an important background for the third subsection. The third subsection describes \emph{attentional RNN Encoder-Decoder}, which is used in our work.

\subsection{Commit Messages}
Our work is motivated by the findings of the studies by Buse \emph{et al.}~\cite{Buse:2010:ADP:1858996.1859005} and by Jiang and McMillan~\cite{jiang2017icpc-era}. The results of the two studies indicate three things. First, commit messages are pervasive and desired. Buse \emph{et al.} examined 1k commits from five mature software projects and found that 99.1\% of the commits have non-empty messages. Jiang and McMillan collected over 2M commit messages from 1k projects. 


Second, human-written commit messages are short. In Buse \emph{et al.}'s study, the average size of the 991 non-empty commit messages is 1.1 lines. Similarly, the study of Jiang and McMillan shows that 82\% of the commit messages have only one sentence. 


Third, commit messages contain various types of information not solely summaries of code changes. Buse \emph{et al.} manually analyzed 375 commit messages and found that the messages are not only about what the changes are but also about why the changes are made. Supported by the three findings, our technique aims to generate one-sentence commit messages which mimic the human-written commit messages.



\subsection{RNN Encoder-Decoder Model}
Neural Machine Translation (NMT) is neural networks that model the translation process from a source language sequence $x=(x_1, ..., x_n)$ to a target language sequence $y=(y_1, ..., y_n)$ with the conditional probability $p(y|x)$~\cite{Alexandru2017ICPC-ERA,Luong2015CoRR}. Cho \emph{et al.} introduced \emph{RNN Encoder-Decoder} as an NMT model~\cite{Cho2014CoRR}, which is commonly used and can produce state of the art translation performance~\cite{Sennrich2016WMT,Luong2016CoRR}. As a promising deep learning model, \emph{RNN Encoder-Decoder} has been used in addressing other software engineering tasks~\cite{Gu2016FSE,Alexandru2017ICPC-ERA}.

\emph{RNN Encoder-Decoder} has two recurrent neural networks (RNNs). One RNN is used to transform source language sequences into vector representations. This RNN is called the encoder. The other RNN is used to  transform the vector representations to the target language sequences, which is called the decoder. 

\subsubsection{Encoder}
The input of the encoder is a variable-length sequence $x=(x_1, ..., x_T)$. The encoder takes one symbol at a time as shown in Figure~\ref{fig:encoder}. As an RNN, the encoder has a hidden state $h$, which is a fixed-length vector. At a time step $t$, the encoder computes the hidden state $h_t$ by:
\begin{equation}
\label{eq:encoder}
h_t = f(h_{t-1}, x_t)
\end{equation}
where $f$ is a non-linear function. Two common options for $f$ are long short-term memory (LSTM)~\cite{Hochreiter1997} and the gated recurrent unit (GRU)~\cite{Cho2014CoRR} (due to space limit, we do not describe these two unit types in detail here). For example, Bahdanau \emph{et al.} use GRU~\cite{Bahdanau2015CoRR} and Sutskever \emph{et al.} use LSTM~\cite{Sutskever2014NIPS}.  The last symbol of $x$ should be an end-of-sequence ({\small \texttt{<eos>}}) symbol which notifies the encoder to stop and output the final hidden state $h_T$, which is used as a vector representation of $x$.

\subsubsection{Decoder}
\label{subsubsec:decoder}
\begin{figure}[!tb]
\centering
\begin{minipage}[b]{1.5in}
	\centering
	\includegraphics[width=0.75in]{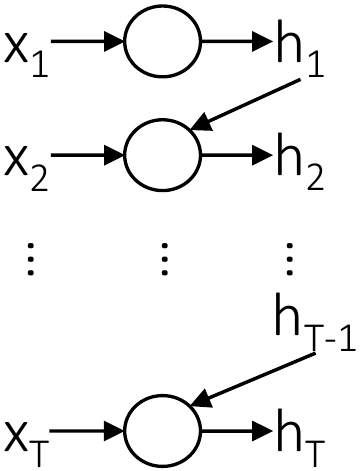}
	\caption{The architecture of the encoder in \emph{RNN Encoder-Decoder}}
	\label{fig:encoder}
\end{minipage} \qquad
\begin{minipage}[b]{1.5in}
	\centering
	\includegraphics[width=1.1in]{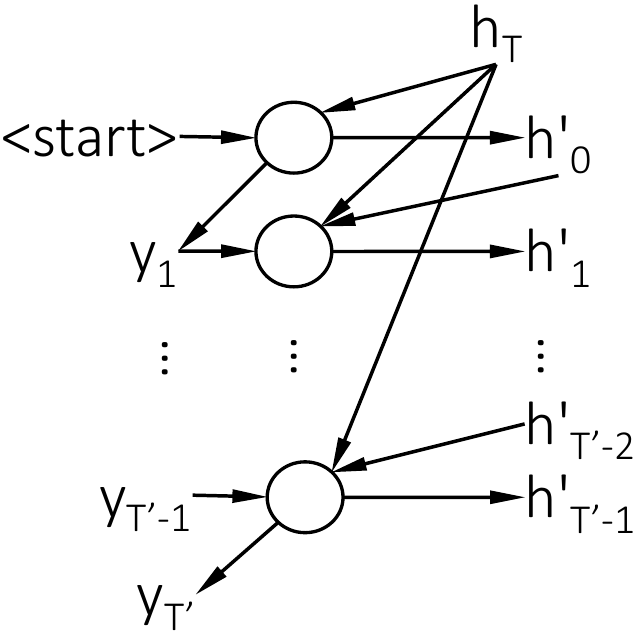}
	\caption{The architecture of the decoder in \emph{RNN Encoder-Decoder}}
	\label{fig:decoder}
\end{minipage}
\end{figure}

Figure~\ref{fig:decoder} shows the RNN of the decoder. The output of the decoder is the target sequence $y=(y_1, ..., y_{T'})$. One input of the decoder is a {\small \texttt{<start>}} symbol denoting the beginning of the target sequence. At a time step $t$, the decoder computes the hidden state $h'_t$ and the conditional distribution of the next symbol $y_t$ by:
\begin{equation}
\label{eq:decoder}
h'_t = f(h'_{t-1}, y_{t-1}, h_{T})
\end{equation}
\begin{equation}
\label{eq:decoder_py}
p(y_t|y_{t-1}, ..., y_1, h_{T}) = g(h'_t, y_{t-1}, h_{T})
\end{equation}
where $h_T$ (generated by the encoder) is called the \emph{context vector}; $f$ and $g$ are non-linear functions. Function $f$ here and $f$ in Equation~\ref{eq:encoder} are often the same. Function $g$ must produce valid probabilities. For example, softmax can be used as $g$. The decoder finishes when it predicts an {\small \texttt{<eos>}} symbol.
 
\subsubsection{Training Goal}
The encoder and the decoder are jointly trained to maximize the conditional log-likelihood: 
\begin{equation}
\max_\theta \frac{1}{N}\sum_{i=1}^{N} \log p (y_i|x_i; \theta)
\end{equation}
where $\theta$ is the set of the model parameters; $N$ is the size of the training set; and each $(x_i, y_i)$ is a pair of a source sequence and a target sequence in the training set.

\subsection{Attentional RNN Encoder-Decoder and Nematus}
\label{subsec:attention}
Bahdanau \emph{et al.} introduced the \emph{attentional RNN Encoder-Decoder}, in which attention mechanism is introduced to deal with long source sequences~\cite{Bahdanau2015CoRR}. We use this mechanism in our work because our source sequences, {\diff}s, are much longer than natural language sentences. The attention mechanism includes several modifications in both the encoder and the decoder, which we describe in the following subsections.


\subsubsection{Encoder}
The encoder in the attentional model is a bidirectional RNN, which has two RNNs: forward and backward. The two RNNs have the same architecture. The forward RNN is the same as the RNN in the original \emph{RNN Encoder-Decoder} model (Figure~\ref{fig:encoder}), which reads the source sequence $x$ as it is ordered, from $x_1$ to $x_T$. The forward RNN generates a sequence of the hidden states {\small $(\overrightarrow{h}_1, ... \overrightarrow{h}_T)$}. In contrast, the backward RNN reads $x$ in the reversed order, and generates a sequence of the hidden states {\small $(\overleftarrow{h}_T, ... \overleftarrow{h}_1)$}.

In the end, for each symbol $x_i$ in $x$, the encoder outputs $h_i=[\overrightarrow{h_i};\overleftarrow{h_i}]$, which is a concatenation of $\overrightarrow{h_i}$ and $\overleftarrow{h_i}$.

\subsubsection{Decoder}
The decoder computes the hidden state $h'_t$ and the conditional distribution of the next symbol $y_t$ by: 
\begin{equation}
h'_t = f(h'_{t-1}, y_{t-1}, c_t)
\end{equation}
\begin{equation}
p(y_t|y_{t-1}, ..., y_1, c_t) = g(h'_t, y_{t-1}, c_t)
\end{equation}
where $f$ and $g$ are non-linear functions like $f$ and $g$ in Equations~\ref{eq:decoder} and~\ref{eq:decoder_py}. $c_t$ is the distinct context vector for $y_t$, and can be computed by
\begin{equation}
c_t=\sum_{i=1}^{T} \alpha_{ti} h_i
\end{equation}
where $T$ is the length of the input sequence; the weight $\alpha_{ti}$ can be trained jointly with the other components in the model, and $h_i$ is generated by the encoder. Since $c_t$ is designed to introduce the context's impact to $y_t$, \emph{attentional RNN Encoder-Decoder} works better on long source sequences. Therefore, we use this NMT model in this paper rather than the original one. 

\section{Approach}
\label{sec:approach}

\begin{figure}[!b]
\centering
\includegraphics[width=3.2in]{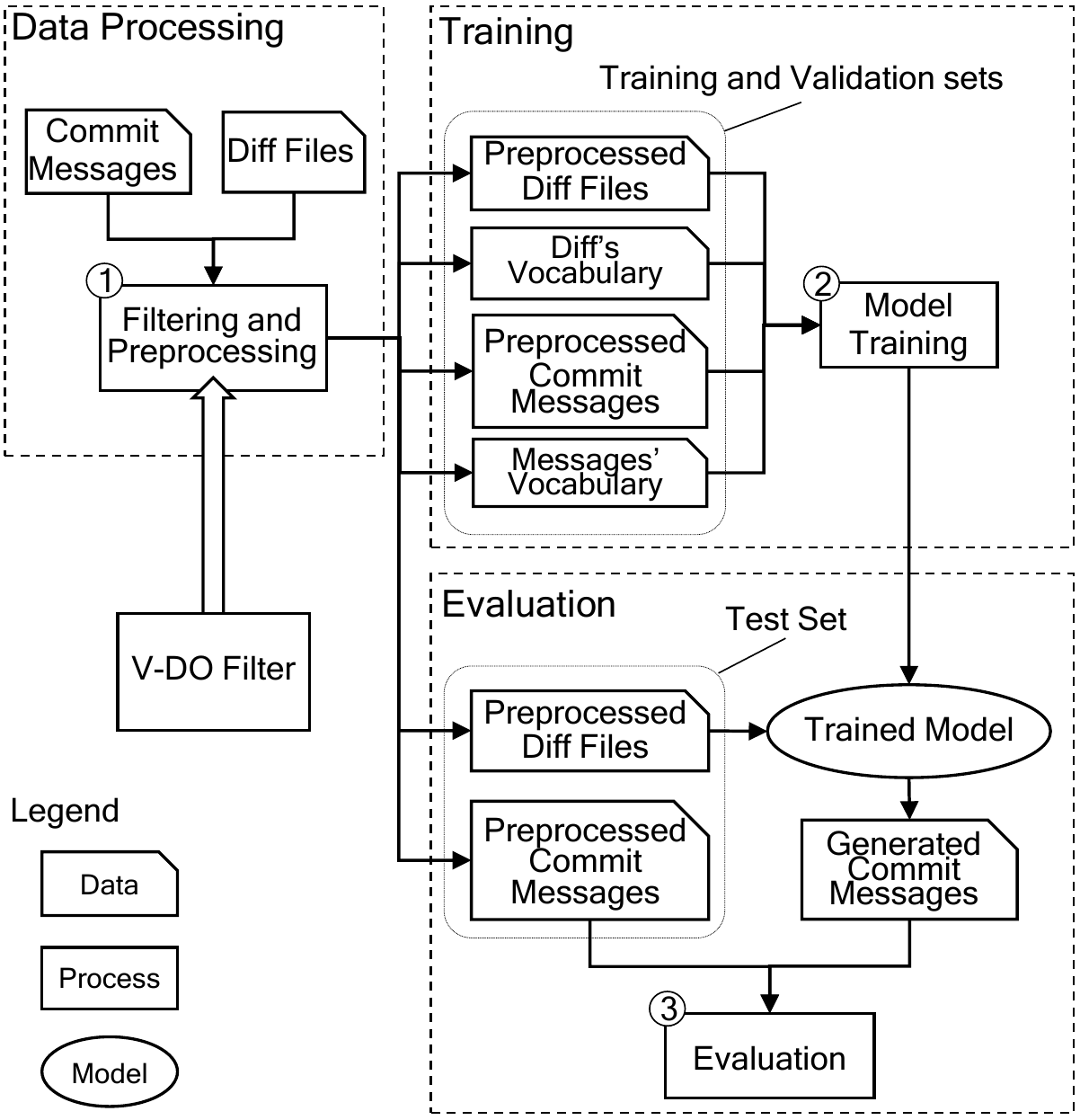}
\caption{The detailed process in Part A, Figure~\ref{fig:paper_overview}. There are three main steps: 1) filtering and preprocessing the data; 2) training a Neural Machine Translation model; 3) evaluating the model, which is Part B, Figure~\ref{fig:paper_overview}.}
\label{fig:a_overview}
\end{figure}

This section describes our approach, including the data set preparation and the NMT training procedure.  This section corresponds to Part A in the paper overview Figure~\ref{fig:paper_overview}, and is detailed in Figure~\ref{fig:a_overview}.

\subsection{Preparing a Data Set for NMT}
\label{sec:part_a_prepare}
We used the commit data set provided by Jiang and McMillan~\cite{jiang2017icpc-era}, which contains 2M commits. The data set includes commits from top 1k Java projects (ordered by the number of stars) in Github. We describe how we prepared the data set for NMT algorithms as follows. 

\subsubsection{Preprocessing the Data Set}
\label{subsec:preprocess}
First, we extracted the first sentences from the commit messages. We used the first sentences as the target sequences because the first sentences often are the summaries of the entire commit messages. Similarly, Gu \emph{et al.} used the first sentences of the API comments as their target sequences~\cite{Gu2016FSE}. Second, we removed issue ids from the extracted sentences and removed commit ids from the {\diff}s, because issue ids and commit ids are unique ids and increase the vocabularies of the source and the target languages dramatically, which in turn cause large memory use of NMT. 



Third, we removed merge and rollback commits (the same practice done by Jiang and McMillan~\cite{jiang2017icpc-era}). Merges and rollbacks are removed because the {\diff}s of merges and rollbacks are often more than thousands of lines, which NMT is not suitable to translate. For the same reason, we also removed any {\diff} that is larger than 1MB.


After the above steps, we have 1.8M commits remaining. Finally, we tokenized the extracted sentences and the {\diff}s by white spaces and punctuations. We did not split CamelCase so that identifiers (e.g., class names or method names) are treated as individual words in this study.  

\subsubsection{Setting Maximum Sequence Lengths for NMT Training}
\label{subsec:maxlength}
A maximum sequence length for both source and target sequences need to be set for an \emph{RNN Encoder-Decoder}~\cite{Bahdanau2015CoRR,Nematus2017}. Since NMT is for translating natural language sentences, maximum sequence lengths for both source and target sequences are often set between 50 to 100~\cite{Bahdanau2015CoRR,Nematus2017}. Because the lengths of our source and target sequences are very different, we set the maximum sequence lengths separately. 


For our target sequences, we set the maximum length at 30 tokens (including words and punctuations), because the first sentences from the commit messages tend to be short. In our data set, 98\% of the first sentences have less than 30 tokens.

For our source sequences, we set the maximum length at 100 tokens because 100 is the largest maximum length used by NMT in natural language translation. Many configurations are possible, and optimizing the maximum {\diff} length for generating commit messages is an area of future work. In pilot studies, a maximum length of 100 outperformed lengths of 50 and 200.



After applying the maximum lengths for source and target sequences (30 and 100), we have 75k commits remaining. 

\subsubsection{V-DO Filter}
\label{subsec:v-do}
We introduced \emph{Verb-Direct Object} (V-DO) filter because we found that the existing messages have different writing styles and some of the messages are poor written, which may affect the performance of NMT.


To obtain a set of commit messages that are in a similar format, we filtered the messages for verb/direct-object pattern. We chose this pattern because a previous study shows that 47\% of commit messages follow this pattern~\cite{jiang2017icpc-era}. To find the pattern, we used a Natural Language Processing (NLP) tool, Stanford CoreNLP~\cite{StanfordNLP}, to annotate the sentences with grammar dependencies. Grammar dependencies are a set of dependencies between parts of a sentences. Considering a phrase, ``program a game'', this phrase has a dependency, which is called ``dobj'' in Stanford CoreNLP, where the governor is ``program'' and the dependent is ``game''. For V-DO filter, we look for ``dobj'' dependencies which represent the verb/direct-object pattern. 

For each sentence, we checked whether the sentence is begun with a ``dobj'' dependency. If the sentence is begun with a ``dobj'', we mark the sentence as a ``dobj'' sentence. In the end, we have 32k commit messages that are ``dobj'' sentences. 


\subsubsection{Generating Training/Validation/Test Sets}
\label{subsec:train_set_and_vocab}
We randomly selected 3k commits for testing, 3k commits for validation, and the rest 26k commits for training. 

\subsubsection{Selecting Vocabularies}
NMT needs predefined vocabularies for commit messages and {\diff}s. 
In the training set,  the commit messages have 16k distinct tokens (words or punctuations) and the {\diff}s have 65k distinct tokens. We selected all the 16k tokens in the commit messages to be the vocabulary of commit messages. We used the most frequent 50k tokens in the {\diff}s to be the vocabulary of {\diff}s. All the tokens that are not in the {\diff} vocabulary only occur once in the training set. 
Additionally, the vocabulary size of 50k is often used by other NMT models~\cite{Luong2016CoRR}.

\subsection{NMT Training and Testing}
\label{sec:nmt_training}

In this section, we describe how we trained and tested an NMT model for generating commit messages. 

\subsubsection{Model}
We used Nematus~\cite{nematus} in our work because it is robust, easy to use, and produced best constrained systems for seven translation directions (e.g., English to German, etc.) in WMT 2016 shared news translation task~\cite{Sennrich2016WMT}. Nematus is based on Theano~\cite{Theano}, and implements the \emph{attentional RNN encoder-decoder} (see Section~\ref{subsec:attention}) with several implementation differences~\cite{nematus}. 

\subsubsection{Training Setting}
We borrowed the training setting that Sennrich \emph{et al.} used to produce the best translation systems in WMT 2016~\cite{Sennrich2016WMT}. The training goal is cross-entropy minimization~\cite{rubinstein2013cross}. The learning algorithm is stochastic gradient descent (SGD) with Adadelta~\cite{zeiler2012adadelta}, which automatically adapts the learning rate. The size of minibatches is 80; the size of word embeddings is 512; the size of hidden layers is 1024. For each epoch, the training set is reshuffled. The model is validated every 10k minibatches by BLEU~\cite{Papineni:2002:BMA:1073083.1073135}, which is a commonly used similarity metric for machine translation. The maximum number of epochs is 5k; the maximum number of minibatches is 10M; and early stopping is used~\cite{nematus}. 
During the training, the model is saved every 30k minibatches. So after the training, a list of models are saved and the ensemble results of the last four models are used for evaluation. 

One key difference between our and Sennrich \emph{et al.}'s training processes is that Sennrich \emph{et al.} used maximum sentence length of 50 for all the languages; we used 30 for commit messages and 100 for {\diff}s as explained in Section~\ref{subsec:maxlength}. 

\subsubsection{Training Details}
We trained on the training set of 26k pairs of commit messages and {\diff}s, with a validation set of 3k pairs. We conducted the training on an Nvidia GeForce GTX 1070 with 8GB memory. The learning algorithm stopped at 210k minibatches. Because a model is saved every 30k minibatches, seven models are saved from this training. The training process took 38 hours. 

\subsubsection{Testing Details}
While we describe our evaluation in the next section, certain technical details are relevant here.  We ran Nematus with the last four saved models on the testing set and we obtained the ensemble result. We used the same GPU as we used in training. The testing process took 4.5 minutes. We note that we followed the standard evaluation procedure for NMT and used a test set of 3k~\cite{Luong2016CoRR,Sennrich2016WMT2,Cho2014CoRR}. 

\section{Evaluation Using An Automatic Metric}
\label{sec:eval}
In this section, we evaluate the generated messages from our approach that we described in the last section.  Our objective is to assess the similarity between the generated messages and the reference messages in the test set. This section corresponds to Part B in the paper overview Figure~\ref{fig:paper_overview}. Note that this evaluation is distinct from the experiment with human evaluators that we describe in Section~\ref{sec:human_study}, which is also a component of ``Part B.'' In this section we ask:

\begin{enumerate}
\item[RQ1] Compared to the messages generated by a baseline, are the messages generated by the NMT model more or less similar to the reference messages?
\item[RQ2] Are the messages generated by the NMT model more or less similar to the reference messages when V-DO filter is enabled or disabled?
\end{enumerate}

We ask RQ1 to evaluate the NMT model compared to a baseline, which we describe in the following subsection. We ask RQ2 in order to evaluate the impact of V-DO filter. 
In the following subsections, we first introduce the baseline for RQ1. Then, we introduce the metric for measuring similarity between two messages. Finally, we report our results for the research questions.

\subsection{Baseline: MOSES}
We used MOSES~\cite{koehn2007moses} as the baseline in RQ1. MOSES is a popular statistical machine translation software, which is often used as a baseline in evaluating machine translation systems~\cite{Callison-Burch:2011:FWS:2132960.2132964,koehn-hoang:2007:EMNLP-CoNLL2007}. For example, Iyer \emph{et al.} used MOSES as a baseline when they evaluated Code-NN~\cite{iyer2016acl}. To run MOSES for translating {\diff}s to commit messages, we trained a 3-gram language model using KenLM~\cite{Heafield2011WMT,Heafield-estimate}, which is the same procedure in the study of Iyer \emph{et al.}~\cite{iyer2016acl}. We did not use Code-NN as a baseline, because, in our pilot study of running Code-NN~\cite{iyer2016acl} to generate commit messages, Code-NN did not generate comparable results. A possible reason is that Code-NN needs parsing source sequences and {\diff}s are not suitable for parsing.

\subsection{Similarity Metric: BLEU}
BLEU~\cite{Papineni:2002:BMA:1073083.1073135} is widely used to measure the similarity between two sentences in evaluation of machine translation systems~\cite{2017opennmt,Luong2016CoRR,Ling2016ACL}. Additionally, BLEU is recommended for assessing an entire test set instead of a sentence~\cite{Papineni:2002:BMA:1073083.1073135}. The calculation of BLEU needs the modified n-gram precisions. For any $n$, the modified $n$-gram precision is calculated by:
\begin{equation}
\label{eq:p_n}
p_{n} = \frac{\sum\limits_{(gen,ref) \in test}  \sum\limits_{ngram \in gen} Cnt_{clip}(ngram)} {\sum\limits_{(gen,ref) \in test} \sum\limits_{ngram \in gen}Cnt_{gen}(ngram)}
\end{equation}
\begin{equation}
\begin{split}
\label{eq:cnt_clip}
Cnt_{clip}&(ngram) = \\
& min(Cnt_{gen}(ngram), Cnt_{ref}(ngram))
\end{split}
\end{equation}
where $test$ is the set of pairs of the generated and the reference messages in the test set; $gen$ is the set of distinct $n$-grams in a generated message; $Cnt_{clip}$ is defined in Equation (\ref{eq:cnt_clip}); $Cnt_{gen}$ is the number of occurrences of an n-gram in a generated message; similarly, $Cnt_{ref}$ is the number of the occurrences of an n-gram in a reference message. Then, BLEU is:
\begin{equation}
BLEU=BP\cdot exp(\sum\limits_{n = 1}^{N} \frac{1}{N}log(p_n))
\end{equation}

\begin{equation}
\label{eq:bp}
\begin{aligned}
BP=\left \{ \begin{aligned}
&1 & \textrm{if } c>r \\
&e^{(1-r/c)} & \textrm{if } c\leq r\\
\end{aligned}\right.
\end{aligned}
\end{equation}
where $N$ is the maximum number of grams; $p_n$ is defined in Equation (\ref{eq:p_n}); $BP$ is defined in Equation (\ref{eq:bp}); $r$ is the sum of the lengths of all the reference messages; $c$ is the sum of the lengths of the generated messages. BLEU scores range from 0 to 100 (in percent). The default value of $N$ is 4, which is used in our evaluation and is commonly used in other evaluations~\cite{2017opennmt,Sennrich2016WMT2,Luong2016CoRR,iyer2016acl,Ling2016ACL,Gu2016FSE}.

\subsection{RQ1: Compared to the Baseline}
The first two rows in Table~\ref{tab:bleu} list the BLEU scores of MOSES and the NMT model we trained in Section~\ref{sec:nmt_training}, which we refer to as \emph{NMT1}. The BLEU score of our model is 31.92 while the BLEU score of MOSES is 3.63, so according to the BLEU metric, the messages generated by the NMT model are more similar to the reference messages than the messages generated by the baseline. One key reason that the attentional NMT model outperforms MOSES is that MOSES does not handle well very long source sequences with short target sequences. Particularly, MOSES depends on Giza++~\cite{och03asc} for word alignments between source and target sequences, and Giza++ becomes very inefficient when a source sequence is 9 times longer than the target sequence or vice versa~\cite{moses_manual}. Table~\ref{tab:bleu} shows that the total length of the generated messages (Len\textsubscript{Gen} in Table~\ref{tab:bleu}) of MOSES is much longer than the total length of the reference messages, which may cause the modified n-gram precisions ($p_1$, $p_2$, $p_3$, and $p_4$), of MOSES to be small. 

\begin{figure}[!b]
	\centering
	\includegraphics[width=3.5in]{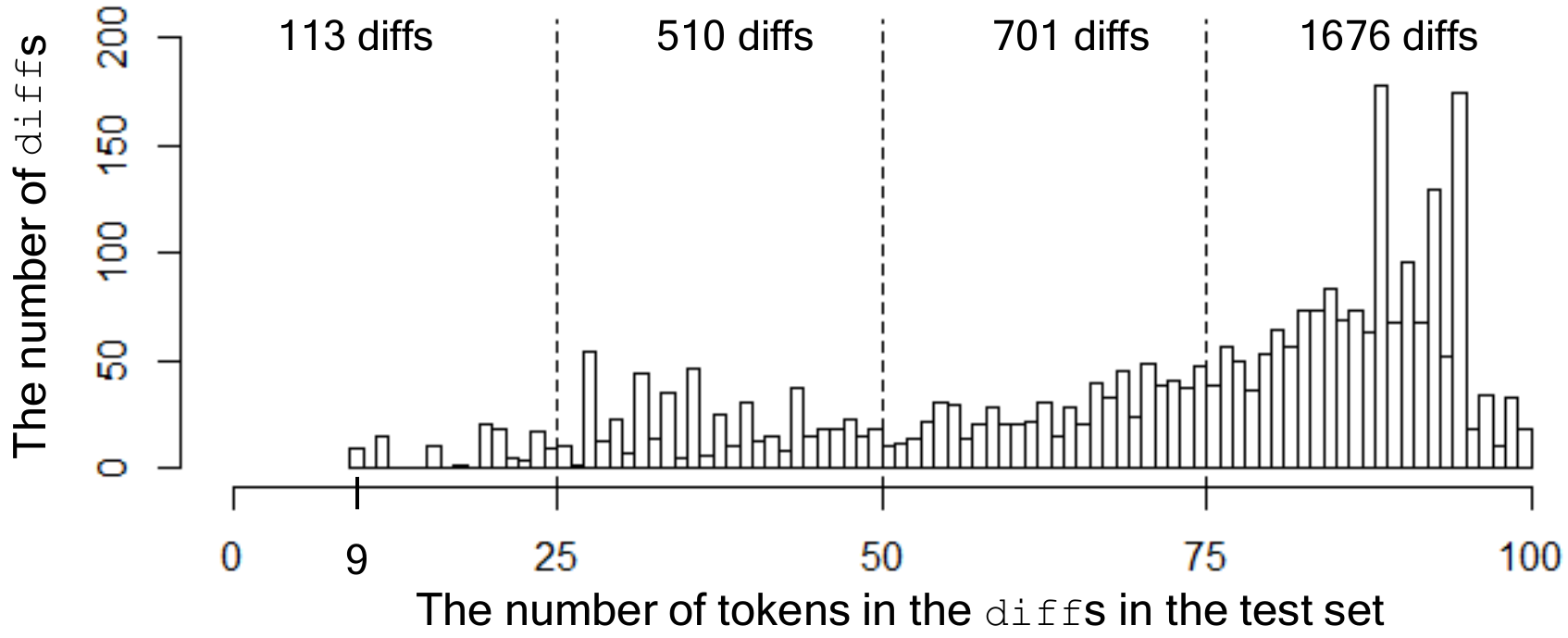}
	\caption{The distribution of the lengths of \texttt{diff}s in the test set}
	\label{fig:test_distribution}
\end{figure}

\begin{figure}[!b]
	\centering
	\includegraphics[width=3.5in]{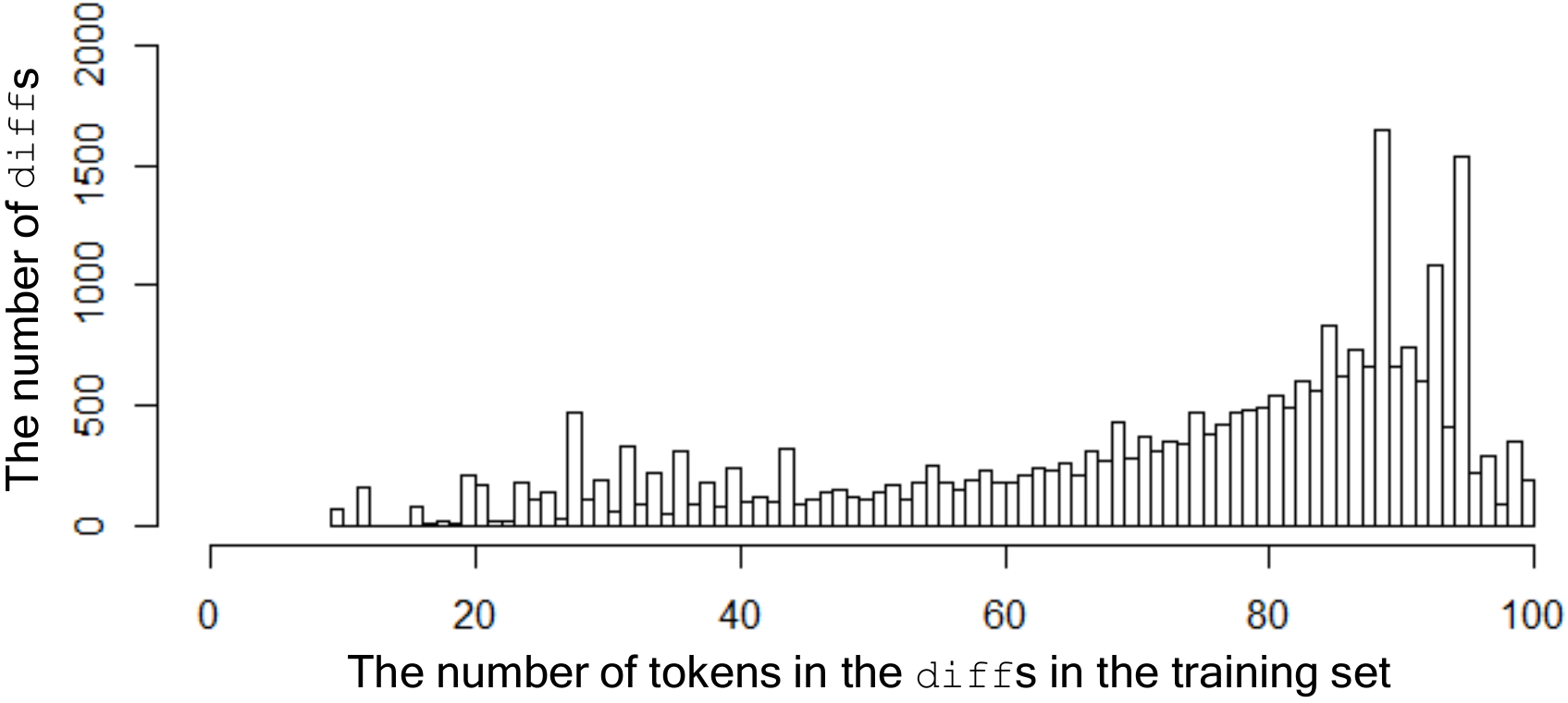}
	\caption{The distribution of the lengths of \texttt{diff}s in the training set}
	\label{fig:train_distribution}
\end{figure}

To further examine the messages generated by our model, we split the test set by the lengths of the {\diff}s into four groups and calculated BLEU scores separately for each group. Figure~\ref{fig:test_distribution} shows the distribution of the lengths of {\diff}s in the test set and Table~\ref{tab:effect_diff_len} shows the BLEU scores for the {\diff}s. This table shows that the {\diff}s that have more than 75 tokens have the highest BLEU score. One possible reason is that there are many more {\diff}s that have more than 75 tokens than the other smaller {\diff}s. Figure~\ref{fig:train_distribution} shows the distribution of the {\diff} lengths in the training set. This figure shows that the training set is populated by larger {\diff}s, which may cause the model to fit the larger {\diff}s better.

In Table~\ref{tab:effect_diff_len}, the modified 4-gram precision, $p_4$, is 7.6 when {\diff} lengths are between 25 and 50, and becomes 42.3 when {\diff} lengths are larger than 75. This increase of $p_4$ means that the number of the 4-grams that are shared by the generated and reference messages increase dramatically when the lengths of {\diff}s increase to more than 75 tokens. In contrast, $p_4$ changes much less (3.1 to 4.5, 4.5 to 7.6) in other cases. 

\begin{table}[!tb]
\caption{BLEU Scores (\%) of MOSES and Our Models on the Test Set}
\label{tab:bleu}
\centering
\begin{tabular}{llllllll}
\hline
Model & BLEU & Len\textsubscript{Gen} & Len\textsubscript{Ref} & $p_{1}$ & $p_{2}$ & $p_{3}$ & $p_{4}$ \\
\hline
MOSES & 3.63 & 129889 & 22872 & 8.3 & 3.6 & 2.7 & 2.1 \\ \hline
NMT1 & 31.92 & 24344 & 22872 & 38.1 & 31.1 & 29.5 & 29.7 \\ \hline
\multirow{2}{*}{NMT2} & 32.81 & 21287 & 22872 & 40.1 & 34.0 & 33.4 & 34.3 \\
                                       & 23.10\textsuperscript{*} & 20303 & 18658 & 30.2 & 23.3 & 20.7 & 19.6 \\
\hline
\multicolumn{8}{p{3.3in}}{MOSES is the baseline model. NMT1 is the NMT model with V-DO filter described in Section~\ref{sec:nmt_training}. NMT2 is a model trained without V-DO filter described in Section~\ref{subsec:rq2}. Len\textsubscript{Gen} is the total length of the generated messages ($c$ in Equation (\ref{eq:bp})). Len\textsubscript{Ref} is the total length of the reference messages ($r$ in Equation (\ref{eq:bp})). The modified n-gram precision $p_{n}$, where $n=1,2,3,4$, is defined in Equation (\ref{eq:p_n}).} \\
\multicolumn{8}{p{3.3in}}{* This BLEU score is calculated on a test set that is not V-DO filtered described in Section~\ref{subsec:rq2}. The other BLEU scores are tested on a V-DO filtered test set described in Section~\ref{subsec:train_set_and_vocab}.}
\end{tabular}
\end{table}

\begin{table}[!t]
\caption{BLEU Scores (\%) on \texttt{diff}s of Different Lengths}
\label{tab:effect_diff_len}
\centering
\begin{tabular}{llllllll}
\hline
Diff Length & BLEU & Len\textsubscript{Gen} & Len\textsubscript{Ref} & $p_{1}$ & $p_{2}$ & $p_{3}$ & $p_{4}$ \\
\hline
$\leq$ 25 & 6.46 & 870 & 655 & 18.6  & 6.9 & 4.3 & 3.1 \\ \hline
$>25$, $\leq 50$ & 9.31 & 3627 & 3371 & 23.1 & 10.8 & 6.6 & 4.5 \\ \hline
$>50$, $\leq 75$ & 12.67 & 4779 & 4418 & 24.8 & 14.1 & 9.8 & 7.6 \\ \hline
$>75$ & 43.33 & 15068 & 14428 & 47.1 & 42.3 & 41.7 & 42.3 \\
\hline
\multicolumn{8}{p{3.3in}}{See Table~\ref{tab:bleu} for explanation of each column name. The BLEU scores are calculated based on the test results generated by Model1, the NMT model with V-DO filter trained in Section~\ref{sec:nmt_training}.}
\end{tabular}
\end{table}


\subsection{RQ2: Impact of V-DO Filter}
\label{subsec:rq2}
Besides NMT1 (the NMT model trained with V-DO filter in Section~\ref{sec:approach}), we trained another model without V-DO filter, which we refer to as \emph{NMT2}. In this subsection, we compare NMT1 and NMT2 to see the impact of V-DO filter.

\subsubsection{Data Set and Training Process for NMT2}
Without V-DO filter, the data set has 75k commits. First, we extracted the test set that is used by NMT1 so that we can compare the test results. Then, from the remaining 72k commits, we randomly selected 3k commits to be another test set, which may contain messages that do not follow the V-DO pattern. We refer to the first test set as \emph{Test1} (with V-DO filter), and the second test set as \emph{Test2} (without V-DO filter).

Then, we randomly selected 3k for validation and used the rest 66k commits for training. We note that the training set of NMT1 has only 26k commits, so NMT2 has 2.5 times more training data than NMT1. The training set includes 45k distinct tokens in commit messages and 110k distinct tokens in {\diff}s. Similar to the vocabulary setting we used in Section~\ref{subsec:train_set_and_vocab}, we used all the 45k tokens to be the vocabulary of commit messages. We used the most frequent 100k tokens in {\diff}s to be the vocabulary of {\diff}s. All the tokens that are not included in the vocabulary only occur once in the training set. We followed the same process described in Section~\ref{sec:nmt_training}. The training process took 41 hours. The testing process for Test1 took 21.5 minutes and Test2 took 20 minutes.

\subsubsection{Results}
The third and fourth rows in Table~\ref{tab:bleu} show the BLEU scores of NMT2 on Test1 and Test2, which are 32.81 and 23.10 respectively. Comparing the BLEU scores of NMT1 and Test1, the result shows that the messages generated by NMT2 are more similar to the reference messages in Test1. This finding indicates that although the training set without V-DO filter has low-quality messages, there are valuable commits that do not follow the V-DO pattern but help the NMT model improve over Test1 which follow the V-DO pattern. 

However, the BLEU score of Test2 is about 10 percent lower than the BLEU score of Test1, which means that NMT2 does not perform well over the commits that do not follow the V-DO pattern. For example, a reference message in Test2 is ``7807cb6 ca7a229'', which should be version numbers. For such reference messages in Test2, the NMT model cannot generate the same version numbers and is not meant to generate such numbers. However, similar messages in the training set cause the NMT model to try to generate such numbers for commit messages. For example, a generated message in Test2 is ``Dd38b1cc2 92007d1d7'' while the reference message is ``Run only on jdk7 for the moment''.

\section{Human Evaluation}
\label{sec:human_study}
In this section, we ask human experts to evaluate the generated messages by the NMT model we described in Section~\ref{sec:approach}. In Section~\ref{sec:eval}, we evaluated our model by the automatic metric, BLEU. Our human study complements the evaluation that uses BLEU in two ways. First, although BLEU is a widely used metric that enables us to compare our model with others and to deliver reproducibility, BLEU is not recommended for evaluating individual sentences~\cite{Papineni:2002:BMA:1073083.1073135}. Our human study can show how our model perform on individual messages. Second, BLEU calculates the textual similarity between the generated and the reference messages, while the human study can evaluate the semantic similarity. 

In this study, we hired 20 participants for 30 minutes each to evaluate the similarity in a survey study. Two participants are computer science Ph.D. students and 18 participants are professional programmers with 2 to 14 years experience. In the rest of this subsection, we describe our survey design, the process of conducting the survey, and the survey results.

\subsection{Survey Design}
\begin{figure}[!t]
\centering
\includegraphics[width=3.5in]{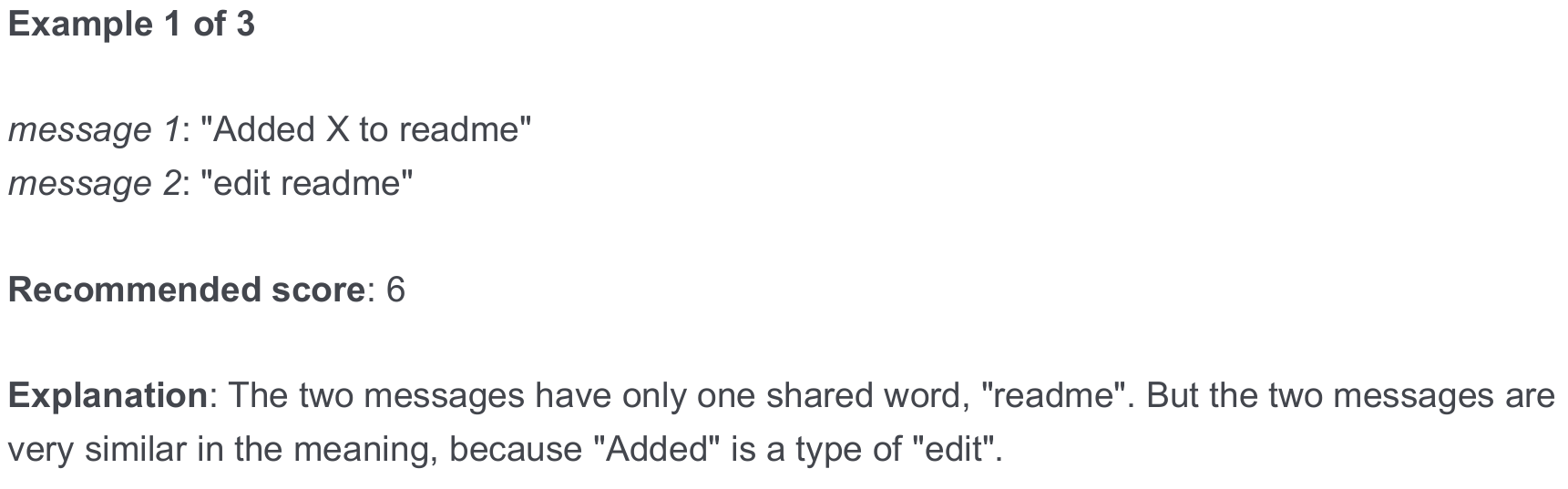}
\caption{An scoring example we gave to the participants in the survey study.}
\label{fig:survey_example1}
\end{figure}

\begin{figure}[!t]
\centering
\includegraphics[width=3.5in]{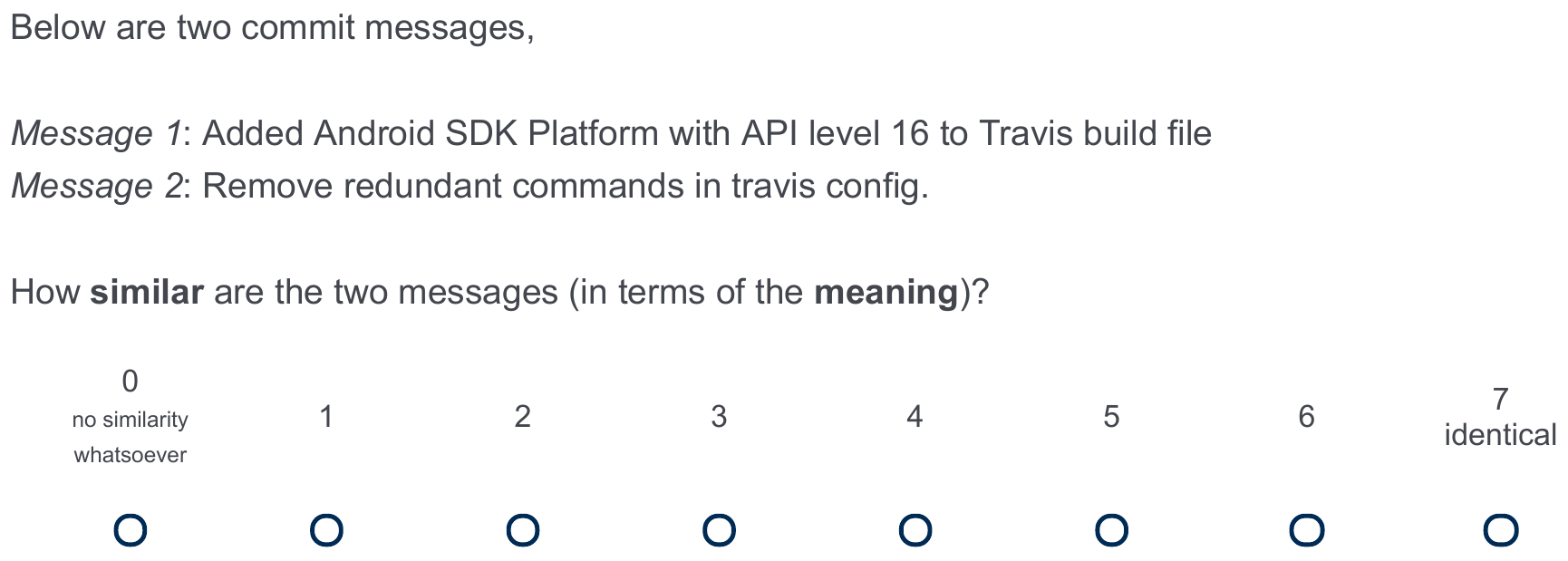}
\caption{One of the pages that we ask the participants to score the similarity. There is an optional text box for the participants to write their justifications in the end of the page. This text box is omitted due to space limit.}
\label{fig:survey_question}
\vspace{-0.4cm}
\end{figure}

We introduce our survey in the first page as: ``This survey will ask you to compare two commit messages by their meaning. You will be able to select a score between 0 to 7, where 0 means there is no similarity and 7 means that two messages are identical.'' We permitted the participants to search the internet for unfamiliar concepts. Then, we gave three scoring examples with recommended scores of 6, 3, and 1. Due to space limit, we present only the first example in Figure~\ref{fig:survey_example1} (all the other examples are available in our online appendix, Section~\ref{subsec:reproducibility}). Then, in the remaining pages of the survey, each page has one pair of the messages, and we asked the participants to score the similarity by meaning. Note that the participants do not know who/what generated the messages. The order of the messages in every page is randomly decided. In the end of the page, there is an optional text box for the participants to enter their justifications. A formal qualitative study about the participants' comments will need to be performed in the future but is beyond the scope of this study. Figure~\ref{fig:survey_question} shows one page of the survey. 


\subsection{Survey Procedure}
First, the pairs of generated/reference messages are randomly ordered in a list. Then, for each participant, a survey is generated with the messages in the list from a given starting point. For example, for the first three participants, the surveys are generated with the messages starting from the first pair in the list. In 30 minutes, the first participant was able to score 107 pairs; the second participant was able to score 61 pairs; the third participant was able to score 99 pairs. So the first 61 pairs of messages were evaluated by three participants. For the fourth participant, we generated a survey starting from the 62th pair and the participant stopped at 99th pair in 30 minutes. So after the first four participants, we have 99 pairs scored by three participants. Although it would be ideal if we obtain three scores for every pair, we did not enforce all the pairs being scored by three participants because we want to have more pairs scored with the limited number of participants. In the end, 226 pairs were scored by three participants, 522 pairs were scored by two participants, and 235 pairs were scored by one participant.

\subsection{Results}
\begin{figure}[!t]
\centering
\includegraphics[width=2.6in]{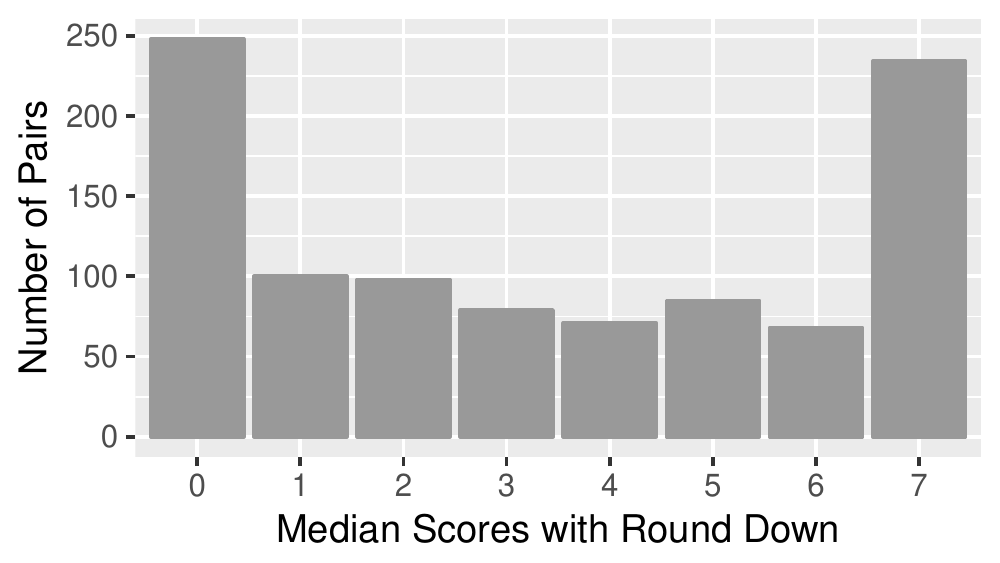}
\caption{The distribution of the median scores obtained in the human study. There are 983 scores in the figure. Each score is the median score of the scores made by one to three human experts for a generated message. The scores range from 0 to 7, where 0 denotes the generated message is not similar to the reference message at all, and 7 denotes the generated message is identical to the reference message. The most frequent scores are 0 and 7. There are 248 messages scored  0 and 234 messages scored 7. For the rest of the scores, the number of messages ranges from 68 to 100.}
\label{fig:survey_result}
\vspace{-0.4cm}
\end{figure}

Figure~\ref{fig:survey_result} shows the distribution of the median scores of the semantic similarity of the generated/reference messages. To be conservative, we round down the median scores. For example, if a generated message has two scores, 1 and 2, and the median score is 1.5, we round down the median score to 1. In total, 983 generated commit messages have scores made by the participants. Zero and seven are the two most frequent scores. There are 248 messages scored 0 and 234 messages scored 7, which shows that the performance of our model tends to be either good or bad. 

\begin{figure}[!b]
	\centering
	\includegraphics[width=2.8in]{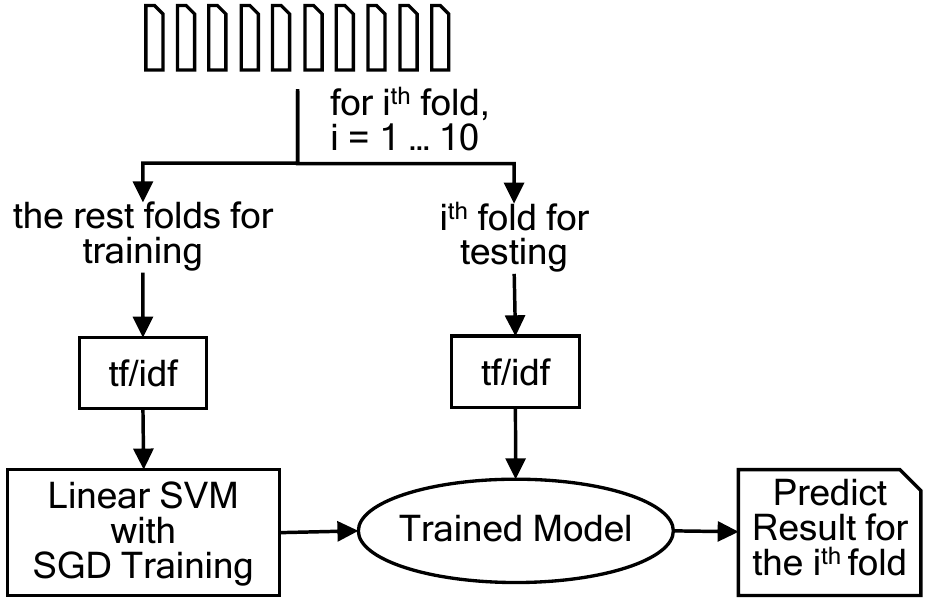}
	\caption{Outline of our cross-validation process.}
	\label{fig:cross_val}
\end{figure}

\section{Quality Assurance Filter}
\label{sec:qa_filter}
Based on the results from our study with human evaluators (Section~\ref{sec:human_study}), we propose a quality assurance filter (QA filter) to automatically detect the {\diff}s for which the NMT model does not generate good commit messages. By building this filter, we investigate whether it is possible to automatically learn the cases where our NMT model does not perform well. In this section, we describe the method of our filter, how we evaluate the filter, and the performance of the filter.  This section corresponds to Part C in the paper overview Figure~\ref{fig:paper_overview}.

\subsection{QA Filter}
Our method of QA filter has three steps. First, we prepared the gold set. We used the evaluated messages and the corresponding {\diff}s in the human study as our gold set. For each {\diff} and the corresponding generated message, there is a score we obtained in the human study (Figure~\ref{fig:survey_result}) that indicates whether the generated message for the {\diff} is similar to the reference message (i.e., the actual human-written message). To be conservative, we labeled the {\diff}s that have scores of zero or one as ``bad'' and all the other {\diff}s as not ``bad''.

Second, we extracted the features of the {\diff}s. We used \emph{term frequency/inverse document frequency} (tf/idf) for every word in a {\diff} as the features. Tf/idf is widely used in machine learning for text processing~\cite{Haiduc2010WCRE}, which is computed based on the frequency of a word in a {\diff} and whether the word is common in the other {\diff}s.

Finally, we used the data set of {\diff}s and their labels to train a linear SVM using stochastic gradient descent (SGD) as the learning algorithm. After we trained the SVM, to predict whether the NMT model will generate a ``bad'' commit message for a {\diff}, we extract tf/idfs from the {\diff} and run the trained SVM with the tf/idfs.

\subsection{Cross-Validation Evaluation}
\begin{figure}[!t]
	\centering
	\includegraphics[width=3.42in]{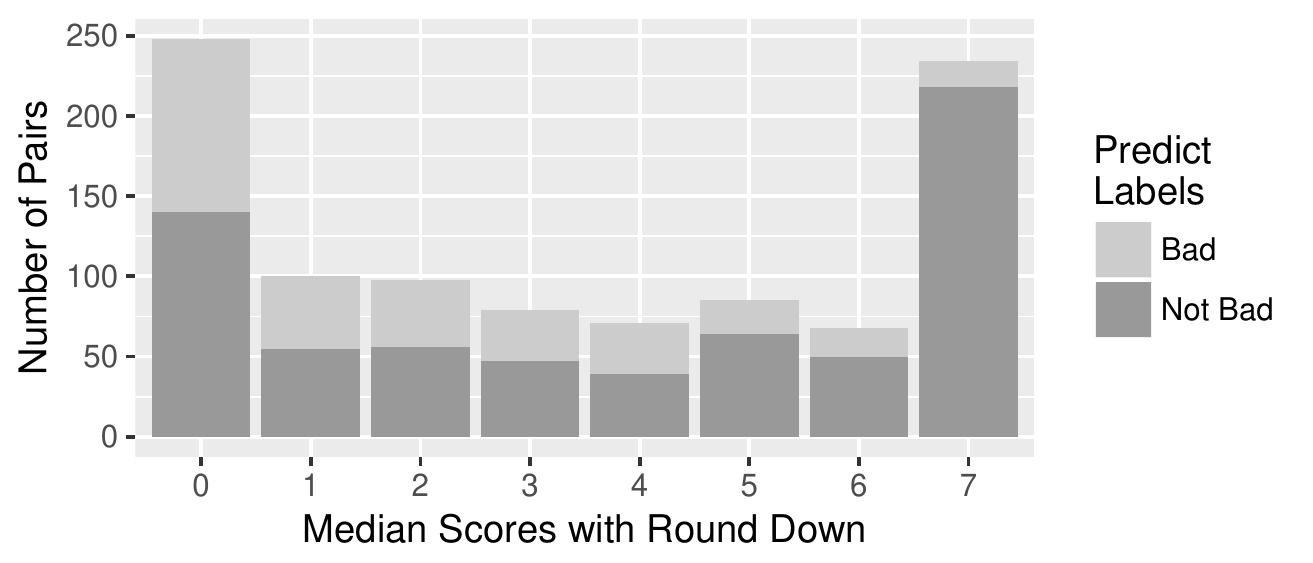}
	\caption{The predict results of the cross evaluation of QA filter. QA filter reduced 108 messages that are scored 0, 45 messages that are scored 1, 42 messages that are scored 2, 32 messages that are scored 3, 32 messages that are scored 4, 21 messages that are scored 5, 18 messages that are scored 6, and 16 messages that are scored 7. We note that although we trained the QA filter with binary labels, ``bad'' and ``not bad'', the evaluation result shows that QA filter is able to reduce more messages for lower scores.}
	\label{fig:qa_result}
	\vspace{-0.4cm}
\end{figure}

Figure~\ref{fig:cross_val} illustrates our 10-fold cross-validation process. We shuffled the gold set first, and split the gold set into 10 folds. For each fold, we trained a SVM model on the other 9 folds, and tested the SVM model on the one fold. In the end, we obtained the test results for every fold. Figure~\ref{fig:qa_result} shows the predicts of all the folds. In terms of detecting {\diff}s for which the NMT model will generate ``bad'' messages, QA filter has 44.9\% precision and 43.8\% recall. Furthermore, if we label the messages with scores of 6 or 7 as ``good'', in this evaluation, QA filter reduced 44\% of the ``bad'' messages at a cost of 11\% of the ``good'' messages.

\section{Example Result}
Table~\ref{tab:example_result} shows a representative example of a generated message that was rated highly by the human experts.  It includes the generated and reference messages, three scores made by three participants, and the corresponding {\diff}.  In this example, the reference message refers to a replacement of a call to a function called \texttt{deactivate()} with a call to a function \texttt{close()}.  To a human reader, that is evident from the \texttt{diff}: a call to \texttt{deactivate()} is removed and a call to \texttt{close()} is added.  The NMT algorithm also picked up on this change, generating text ``Close instead of mCursor.Deactivate.''  

\begin{table}[!h]
	\caption{Example Result}
	\label{tab:example_result}
	\centering
	\begin{tabular}{|l|}
		\hline
		Diff: \rule{0pt}{2.6ex} \\
		{\code{--- a/core/.../CursorToBulkCursorAdaptor.java}} \\
		{\code{+++ b/core/.../CursorToBulkCursorAdaptor.java}} \\
		{\code{@@ -143,8 +143,7 @@ public final class }} \\
		{\code{CursorToBulkCursorAdaptor ... }} \\ 
		{\code{     public void close() \{}} \\
		{\code{         maybeUnregisterObserverProxy();}} \\
		{\code{-        mCursor.deactivate();}} \\
		{\code{- }}\\
		{\code{+        mCursor.close();}} \\
		{\code{     \}}} \\
		{\code{     public int requery(IContentObserver observer, ...}}\\[2pt] \hline
		Generated Message: \rule{0pt}{2.6ex} \\
		``CursorToBulkCursorAdapter . Close must call\\
		mCursor . Close instead of mCursor . Deactivate . '' \\[2pt] \hline
		Reference Message: \rule{0pt}{2.6ex}\\
		``Call   close (  )    instead   of   deactivate (  ) in\\
		CursorToBulkCursorAdaptor . close (  ) '' \\[2pt] \hline
		Scores: 7, 6, 7 \rule{0pt}{2.6ex} \\[2pt] \hline
	\end{tabular}
\vspace{-0.4cm}
\end{table}

\section{Threats to Validity}
One threat to validity is that our approach is experimented on only Java projects in Git repositories, so they may not be representative of all the commits. However, Java is a popular programming language~\cite{tiobe,pypl,stackoverflow}, which is used in a large number of projects. In the future, we will extend our approach to other programming languages.

Another threat to validity is the quality of the commit messages. We collected actual human-written commit messages from Github, and used V-DO filter to obtain a set of relatively good-quality commit messages. But the human-written messages may not contain all the useful information that should be in a commit message. However, our objective in this paper is to generate commit messages that can be learned from the history of the repositories. Further improvement on human-written messages falls outside the scope of this paper.

Another threat to validity is about the human study because of the limited number of the participants. We cannot guarantee that every final score for a generated commit message is fair. We tried to mitigate this threat by hiring as many professional programmers as we can, and having 23\% of the evaluated messages scored by three participants and 53\% of the evaluated messages scored by two participants. 

\section{Discussion and Conclusion}
\label{sec:discussion}
The key advancement that this paper makes to the state-of-the-art is a technique to generate short commit messages that summarize the high-level rationale for a change to software.  As we note in Section~\ref{sec:problem}, we do not claim to be able to provide new insights for completely novel changes to software -- that task is likely to remain in human hands for the foreseeable future.  Instead, we learn from knowledge stored in a repository of changes that have already been described in commit messages.  Several authors in the related literature have observed that many code changes follow similar patterns, and have a similar high-level rationale (e.g., \cite{883028, jiang2017icpc-era}).  Traditionally programmers still need to manually write commit messages from scratch, even in cases where a commit has a rationale that has been described before.  What this paper does is automate writing commit messages based on knowledge in a repository of past changes.

Our strategy was, in a nutshell, to 1) collect a large repository of commits from large projects, 2) filter the commits to ensure relatively high-quality commit messages, and 3) train a Neural Machine Translation algorithm to ``translate'' from \texttt{diff}s to commit messages using the filtered repository.  We then evaluated the generated commit messages in two ways.  First we conducted an automated evaluation using accepted metrics and procedures from the relevant NMT literature (Section~\ref{sec:eval}).  Second, as a verification and for deeper analysis, we also conducted an experiment with human evaluators (Section~\ref{sec:human_study}).

What we discovered is that the NMT algorithm succeeded in identifying cases where the commit had a similar rationale to others in the repository.  The evidence for this is the large bar for item 7 in Figure~\ref{fig:survey_result} -- it means that the human evaluators rated a large number of the generated messages as very closely matching the reference messages.  However, the algorithm also generated substantial noise in the form of low quality messages (note the large bar for item 0).  A likely explanation is that these include the cases that involve new insights which the NMT algorithm is unable to provide.  While creating these new insights from the data is currently beyond the power of existing neural network-based machine learning (a problem observed across application domains~\cite{goodfellow2016deep}), at a minimum we would like to return a warning message to the programmer to indicate that we are unable to generate a message, rather than return a low quality message.  Therefore we created a Quality Assurance filter in Section~\ref{sec:qa_filter}.  This filter helped reduce the number of low quality predictions, as evident in the reduced bar for item 0 in Figure~\ref{fig:qa_result}.

While we do view our work as meaningfully advancing the state-of-the-art, we by no means claim this work is definitive or completed.  We release our complete data set and implementation via an online appendix, noted at the end of Section~\ref{sec:approach}.  Our hope is that other researchers will use this data set and implementation for further research efforts.  Generally speaking, future improvements are likely to lie in targeted training for certain types of commits, combined with detection of change types.  It is probable that very high quality predictions are possible for some types of software changes, but not others.  This work provides a foundation for those and other future developments.

\section{Reproducibility}
\label{subsec:reproducibility}
Our data sets, scripts, and results are accessible via: 

\textbf{\url{https://sjiang1.github.io/commitgen/}} 

\section{Acknowledgments}
This work was partially supported by the NSF CCF-1452959 and CNS-1510329 grants, and the Office of Naval Research grant N000141410037. Any opinions, findings, and conclusions expressed herein are the authors’ and do not necessarily reflect those of the sponsors.

\IEEEpeerreviewmaketitle

\bibliographystyle{abbrv}
\bibliography{main}

\end{document}